\documentclass{pkas}

\def\year{2015} % publication year
\def\month{January} % publication month
\def\volume{00} % journal volume
\def\issue{0} % journal issue
\def\beginpage{1} % first page of article
\def\endpage{4} % last page of article
\def\received{October 31, 2014} % date paper was received by PKAS
\def\accepted{November 30, 2014} % date of acceptance
\date{Received \received ; accepted \accepted}

\newcommand\ion[2]{{#1}\,{\sc #2}} % ions: \ion{C}{iv} = C IV

\def\aj{AJ}
\def\araa{ARA\&A}
\def\apj{ApJ}
\def\apjl{ApJ}

\def\apss{Ap\&SS}
\def\aap{A\&A}

\def\aaps{A\&AS}

\def\mnras{MNRAS}

\def\pasa{Publ.~Astron.~Soc.~Australia}
\def\pasp{PASP}

\def\rmxaa{RMxAA}

\def\arcsec{\hbox{$^{\prime\prime}$}}

\title{
Kinematical Properties of \\Planetary Nebulae with WR-type Nuclei
}

\author[1]{Ashkbiz~Danehkar\thanks{E-mail: ashkbiz.danehkar@students.mq.edu.au}}
\author[2]{Wolfgang~Steffen}
\author[1,3]{Quentin~A.~Parker}
\affil[1]{Department of Physics and Astronomy, Macquarie University, Sydney, NSW 2109, Australia} %; \email{ashkbiz.danehkar@students.mq.edu.au}}
\affil[2]{Instituto de Astronom\'{i}a, Universidad Nacional Aut\'{o}noma de M\'{e}xico, C.P.22860, Ensenada, Mexico} %; \email{wsteffen@astro.unam.mx}}
\affil[3]{Australian Astronomical Observatory, PO Box 915, North Ryde, NSW 1670, Australia}%; \email{quentin.parker@mq.edu.au}}
\def\corrauthor{
A.~Danehkar
}

\def\runningauthor{
A.~Danehkar, W. Steffen \& Q. A. Parker 
}

\def\runningtitle{
Kinematics of WR Planetary Nebulae
}

\def\keywords{
planetary nebulae: general -- ISM: kinematics and dynamics --
stars: Wolf-Rayet
}

\def\abstracttext{
We have carried out integral field unit (IFU) spectroscopy of H$\alpha$, [\ion{N}{ii}] and [\ion{O}{iii}] emission lines for a sample of Galactic planetary nebulae (PNe) with Wolf-Rayet (WR) stars and weak emission-line stars (\textit{wels}). 
Comparing their spatially-resolved kinematic observations with morpho-kinematic models allowed us to
disentangle their three-dimensional gaseous structures.  
Our results indicate that these PNe have axisymmetric morphologies, either bipolar or elliptical. In many cases the associated kinematic maps for the PNe around hot central stars also reveal the presence of so-called fast low-ionization emission regions. %(FLIERs). 
}

\begin{document}
\pkashead %% set title, authors, abstract, etc.

%%%%%%%%%%%%%%%%%%%%%%%%%%%%%%%%%%%%%%%%%%%%%%%%%%%%%%%%%%%%%%%%%%%%%
%%% BEGIN MAIN TEXT HERE %%%%%%%%%%%%%%%%%%%%%%%%%%%%%%%%%%%%%%%%%%%%
%%%%%%%%%%%%%%%%%%%%%%%%%%%%%%%%%%%%%%%%%%%%%%%%%%%%%%%%%%%%%%%%%%%%%

\section{Introduction\label{wc1:sec:introduction}}

A planetary nebula (PN) is a shell of hydrogen-rich material previously ejected by the asymptotic giant branch (AGB) progenitor star during the thermally pulsating phase. Once the central star (CSPN) leaves the AGB phase, the stellar wind contributes some hydrodynamic effects to the ionized shell, and gradually changes its shape. The emission lines emitted by the ionized gas carry important information, which can be used to determine the kinematic features of the nebula. The kinematic analysis of PNe provides some insights into the AGB mass-loss process, the transition from the AGB to the PN phase, and the (pre-)PN evolution \citep[see e.g.][]{Kwok2010}.

The majority of PNe show predominantly axisymmetric morphologies \citep[e.g.][]{Balick1987}, which have introduced considerable problems into theories of their formation and evolution. According to the interacting stellar winds (ISW) theory of nebular formation developed by \citet{Kwok1978}, a slow dense superwind from the AGB phase is swept up by a fast tenuous wind during the PN phase, creating a compressed dense shell. \citet{Kahn1985} extended this model to describe an aspherical mass distribution i.e. highly axisymmetric or bipolar nebulae. This extension later became known as the generalized interacting stellar winds (GISW) theory. But, it is not always consistent with observations and a matter of some controversy \citep[see review by][]{Balick2002}. More complex axisymmetric morphologies recently observed \citep[e.g.][]{Sahai2011} contradict the GISW theory. To accommodate those complexities, \citet{Steffen2013} considered a further generalization of the GISW model incorporating an inhomogeneous AGB wind, and their simulations suggest that bipolar PNe are evolved from rather massive progenitors. A combination of rotating stellar winds and strong toroidal magnetic fields has been also proposed as a mechanism for the equatorial density enhancement and the jet-like outflows \citep[e.g.][]{Garcia-Segura2000,Frank2004}. However, recent MHD simulations showed that single star evolution is unlikely to shape bipolar PNe \citep[][]{Garcia-Segura2014}. It has also been suggested that axisymmetric morphologies can be produced through tidal interaction with a binary partner \citep{Soker2006,Nordhaus2006}. 

The small-scale low-ionization structures (LISs) embedded or not in the global structure has been found in nearly 10\% of Galactic PNe \citep{Gonccalves2009}. Around  50\% of LISs are highly
collimated high-velocity jets  or high-velocity pairs of knots, the so-called fast, low-ionization emission regions \citep[FLIERs;][]{Balick1994}, which have radial velocities of 30--200 km\,s$^{-1}$ with respect to the main bodies. We have not yet understood how the density and velocity structures contrast between the FLIERs and the main body.

This study aims to determine the kinematic structures of a sample of Galactic PNe ionized by Wolf-Rayet (WR) stars and weak emission-line stars (\textit{wels}). 

\begin{figure*}
\begin{center}
{\scriptsize (a) M\,3-30}\\ 
\includegraphics[width=1.1in]{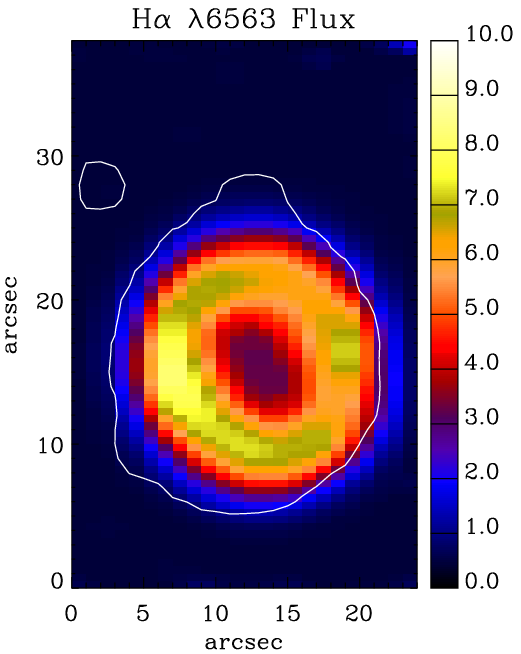}%
\includegraphics[width=1.1in]{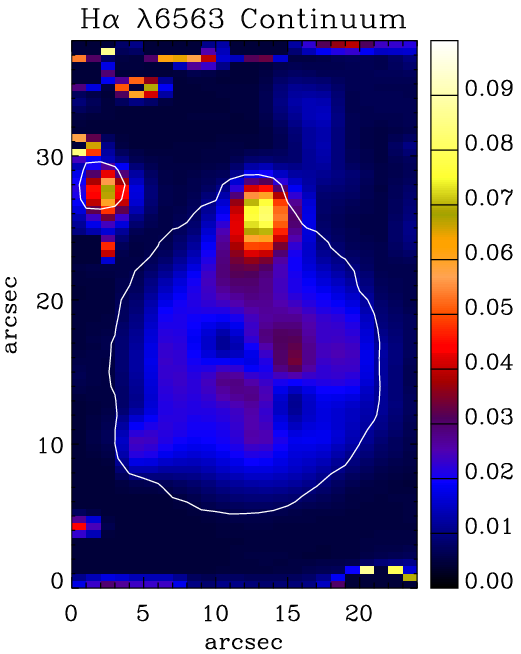}%
\includegraphics[width=1.1in]{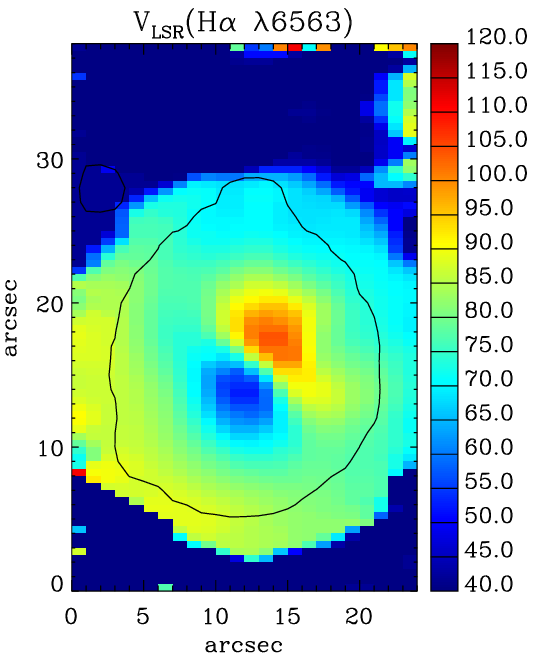}%
\includegraphics[width=1.1in]{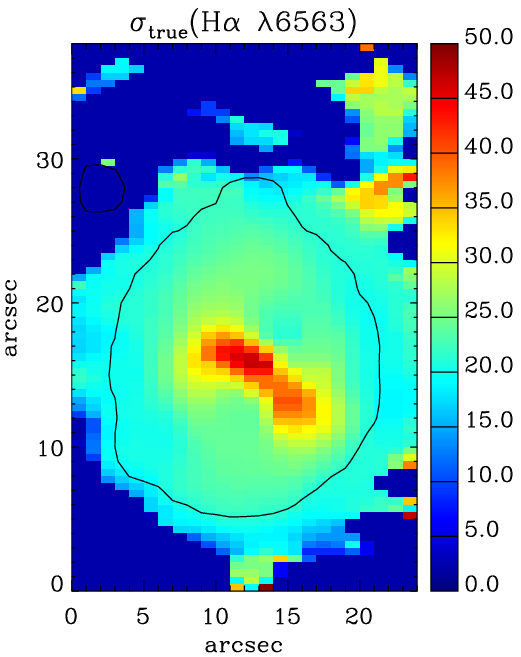}%
\includegraphics[width=1.1in]{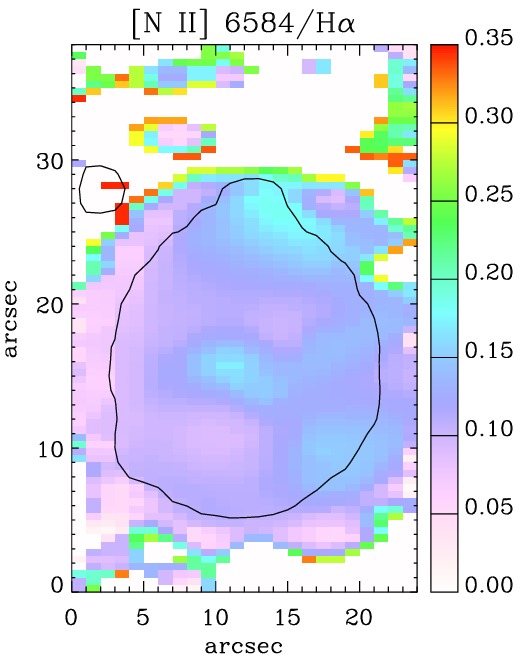}
\includegraphics[width=1.1in]{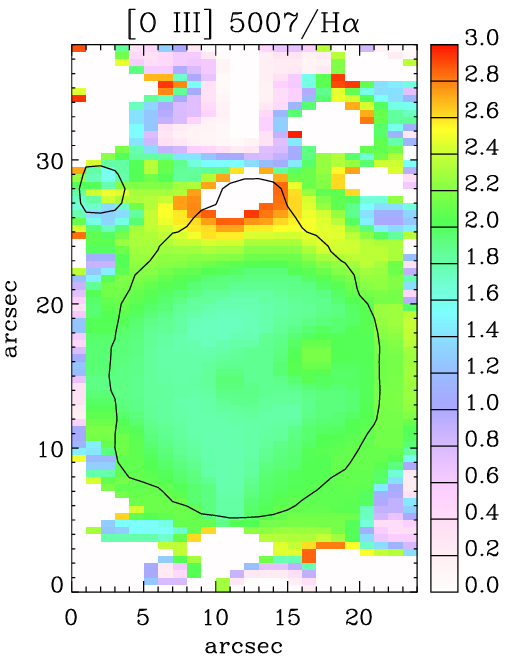}\\
{\scriptsize (b) Hb\,4}\\
\includegraphics[width=1.1in]{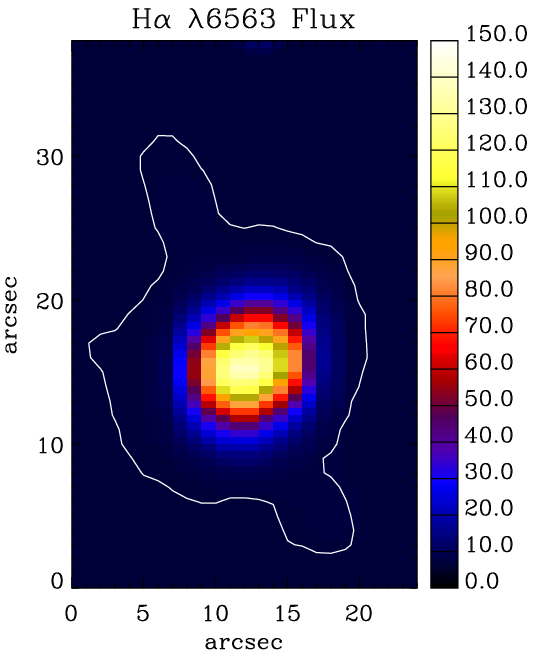}%
\includegraphics[width=1.1in]{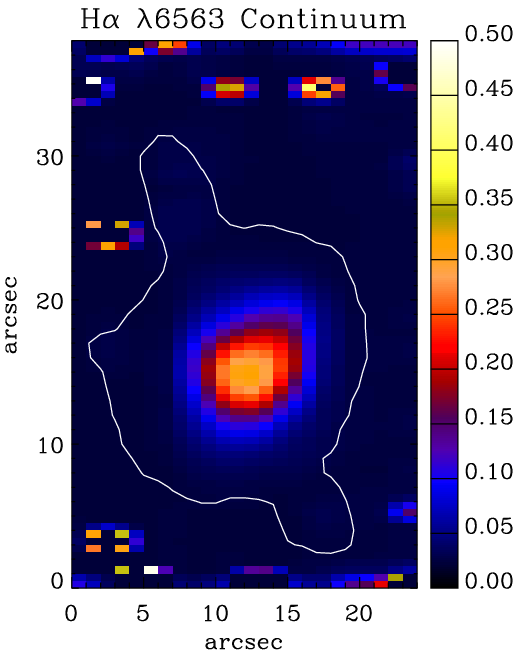}%
\includegraphics[width=1.1in]{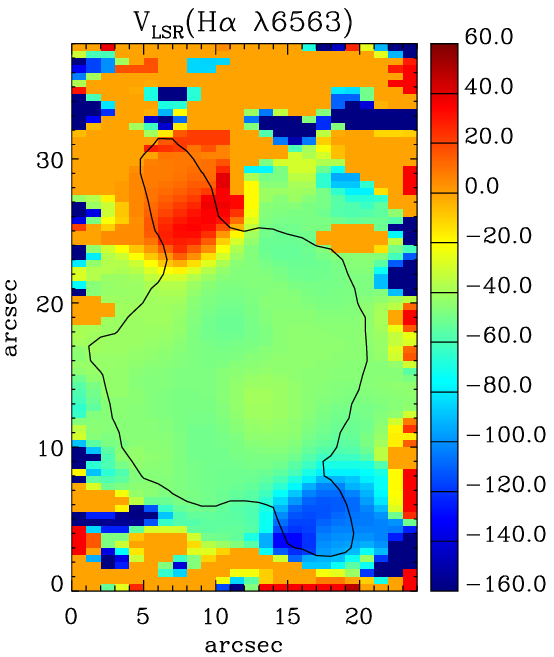}%
\includegraphics[width=1.1in]{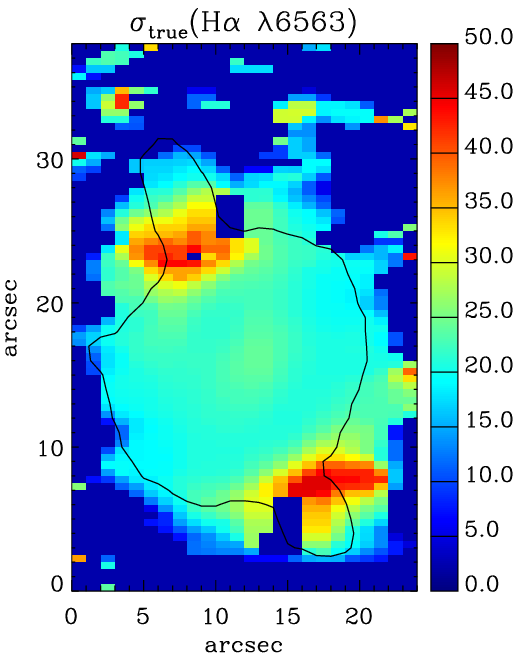}%
\includegraphics[width=1.1in]{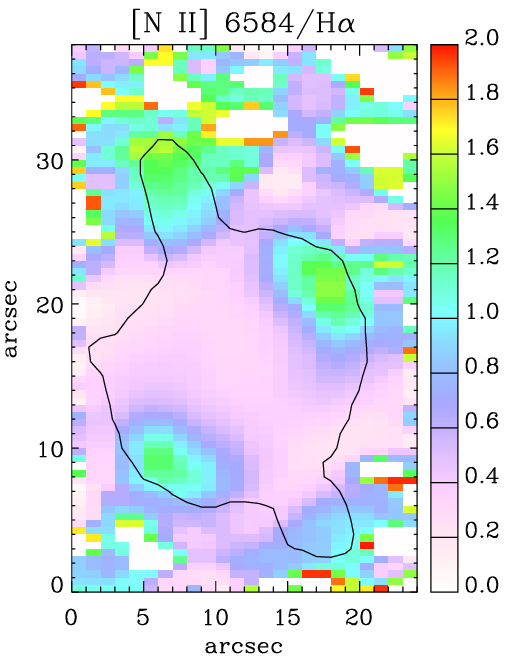}%
\includegraphics[width=1.1in]{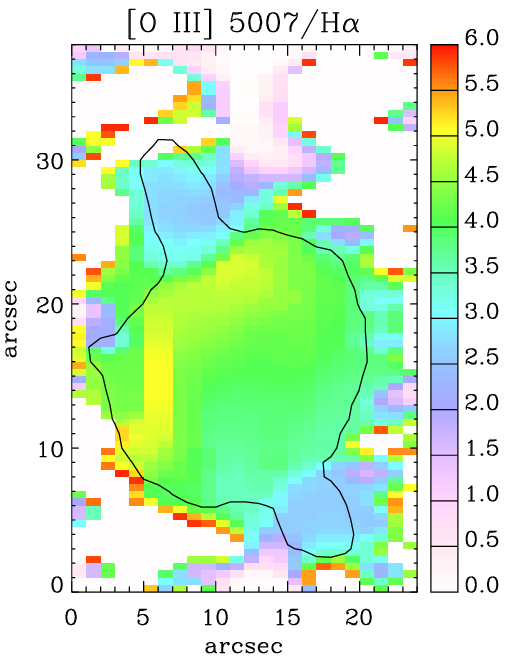}\\
{\scriptsize (c) IC\,1297}\\ 
\includegraphics[width=1.1in]{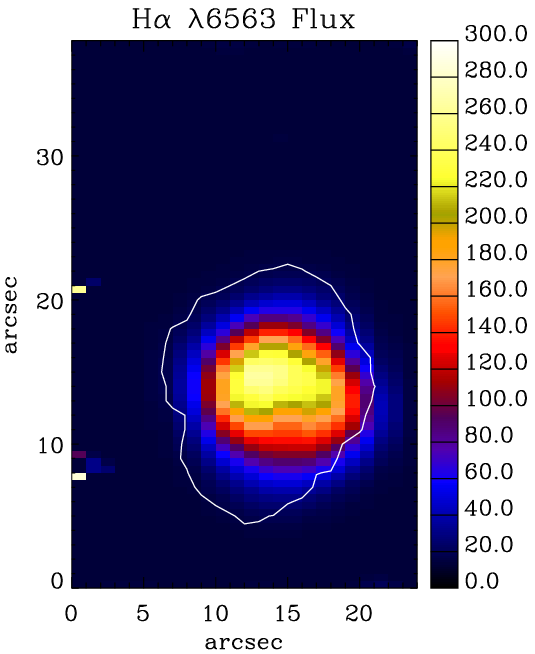}%
\includegraphics[width=1.1in]{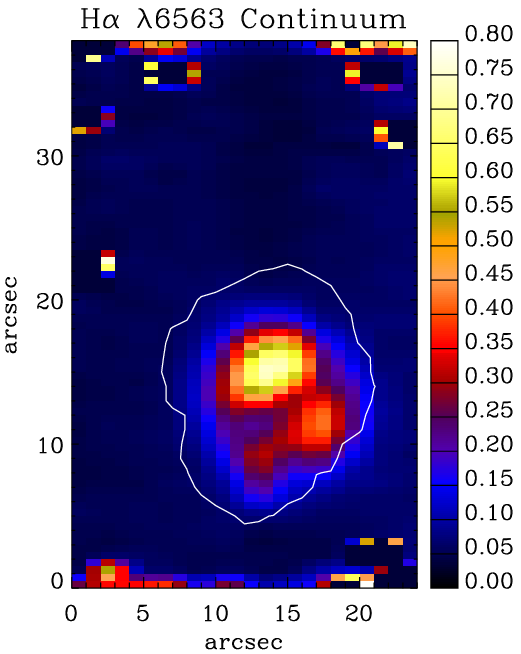}%
\includegraphics[width=1.1in]{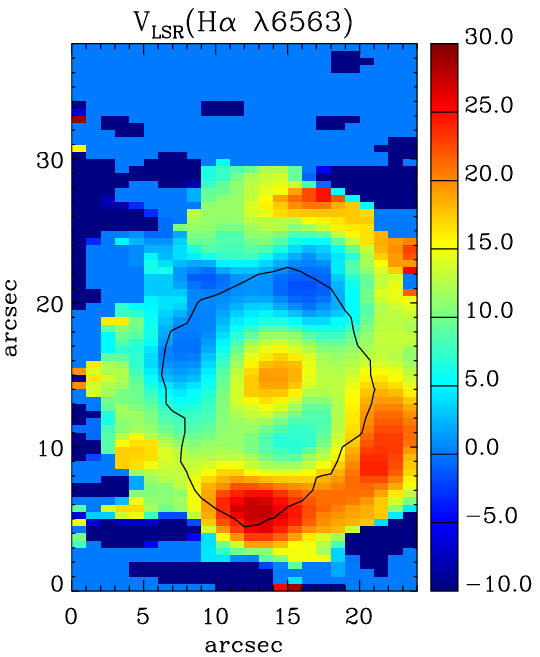}%
\includegraphics[width=1.1in]{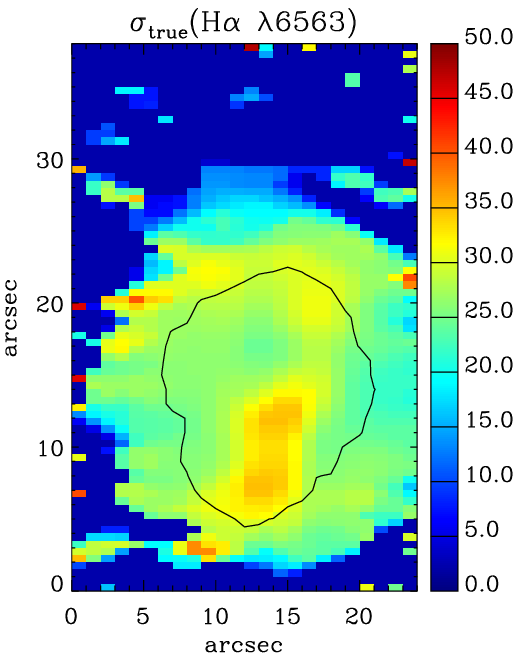}%
\includegraphics[width=1.1in]{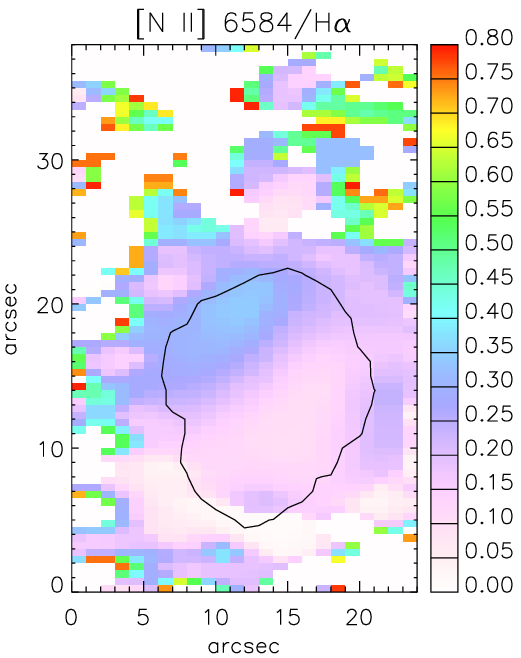}%
\includegraphics[width=1.1in]{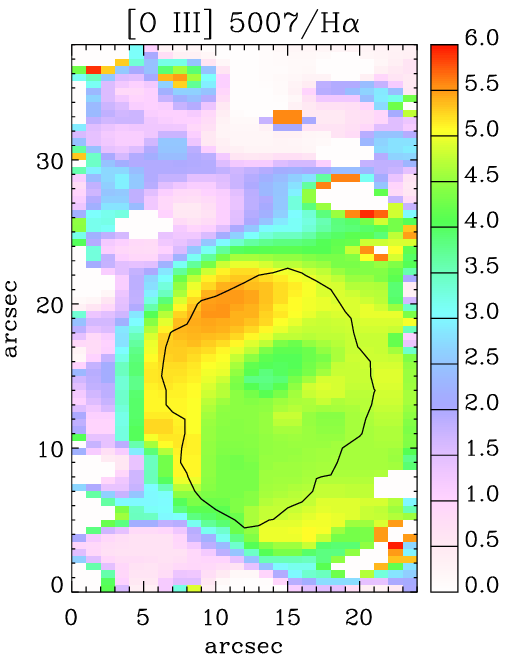}\\
{\scriptsize (d) Th\,2-A}\\ 
\includegraphics[width=1.1in]{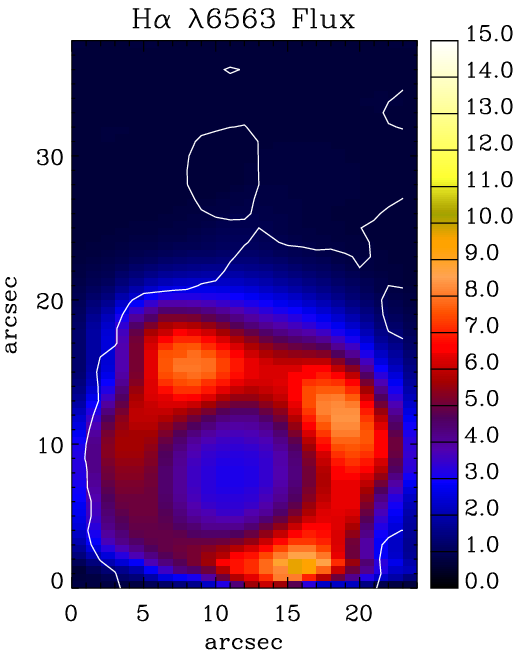}%
\includegraphics[width=1.1in]{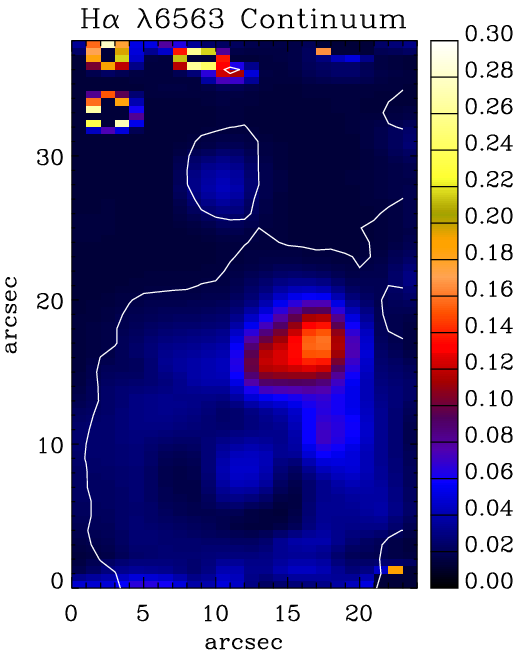}%
\includegraphics[width=1.1in]{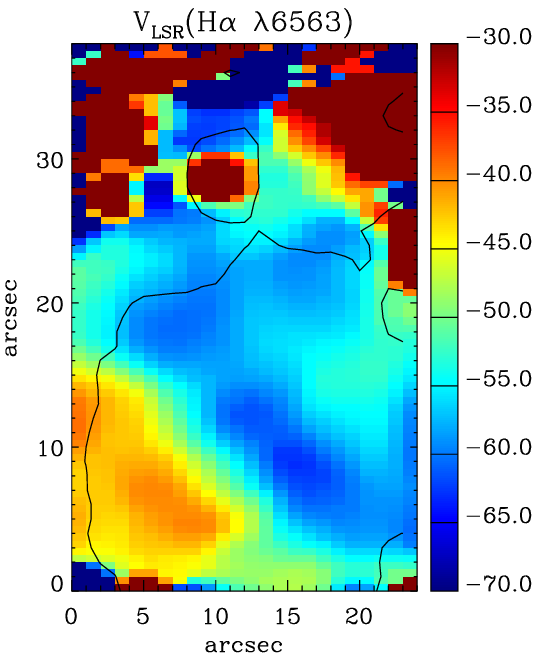}%
\includegraphics[width=1.1in]{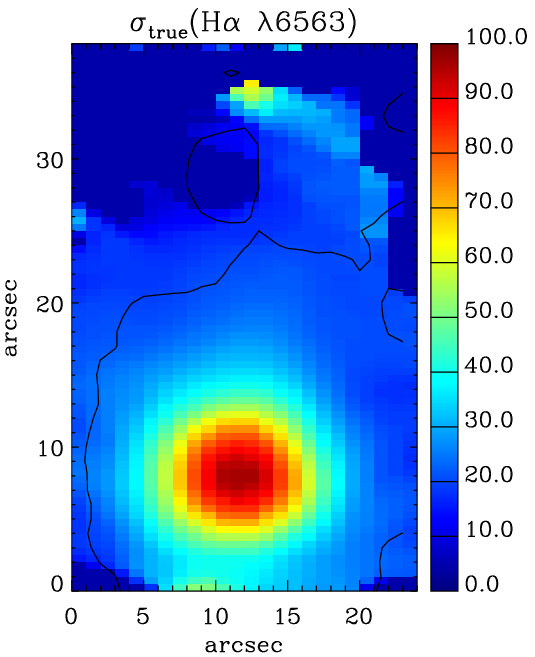}%
\includegraphics[width=1.1in]{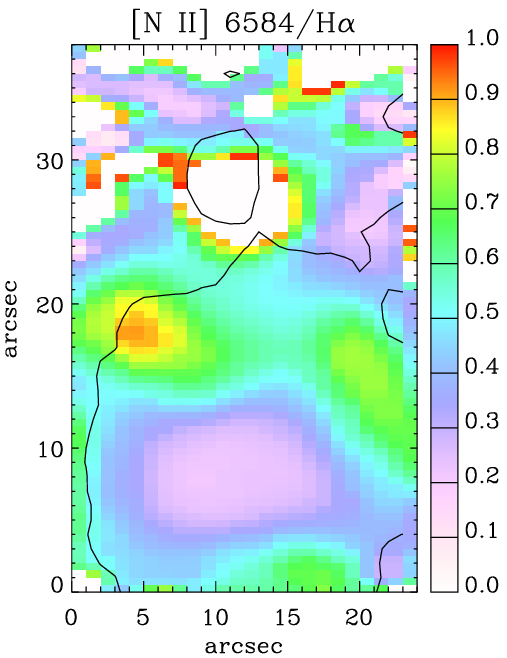}%
\includegraphics[width=1.1in]{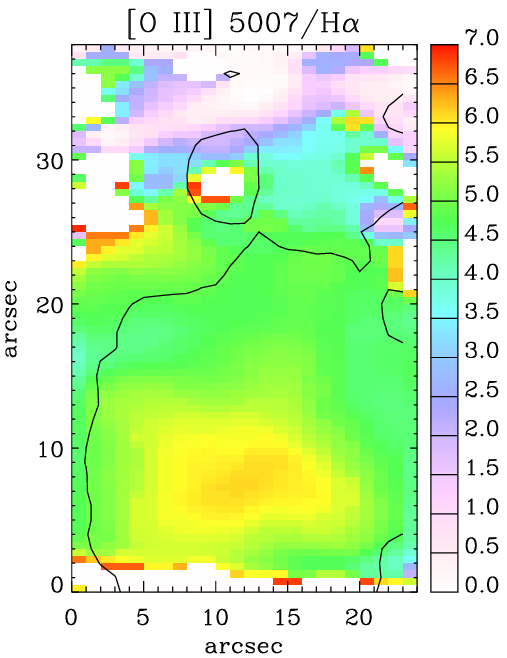}\\
{\scriptsize (e) K\,2-16}\\ 
\includegraphics[width=1.1in]{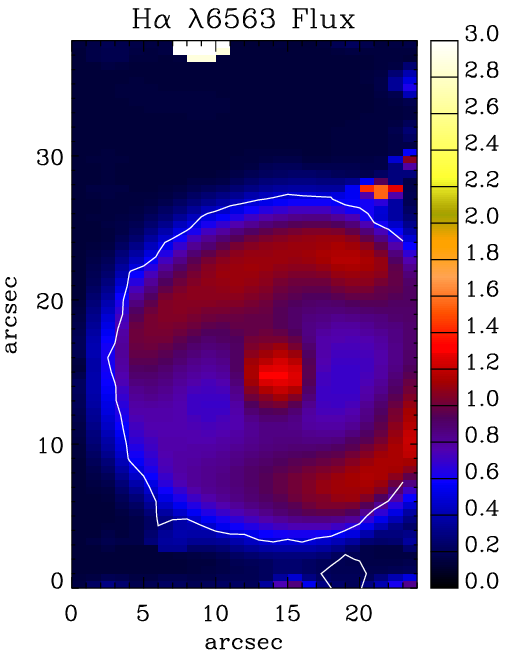}%
\includegraphics[width=1.1in]{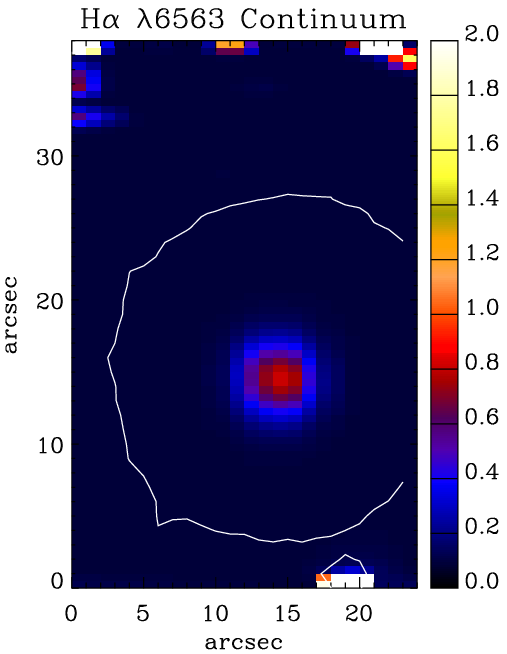}%
\includegraphics[width=1.1in]{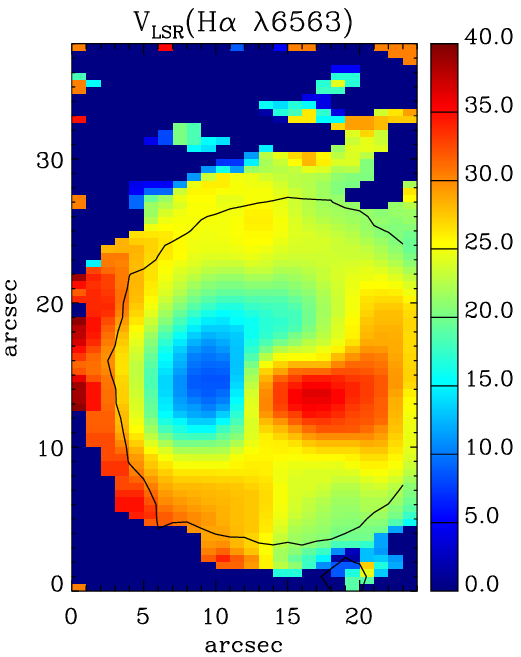}%
\includegraphics[width=1.1in]{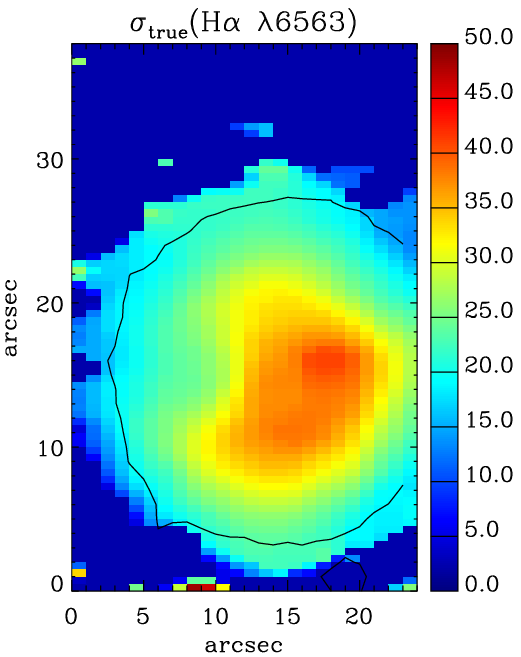}%
\includegraphics[width=1.1in]{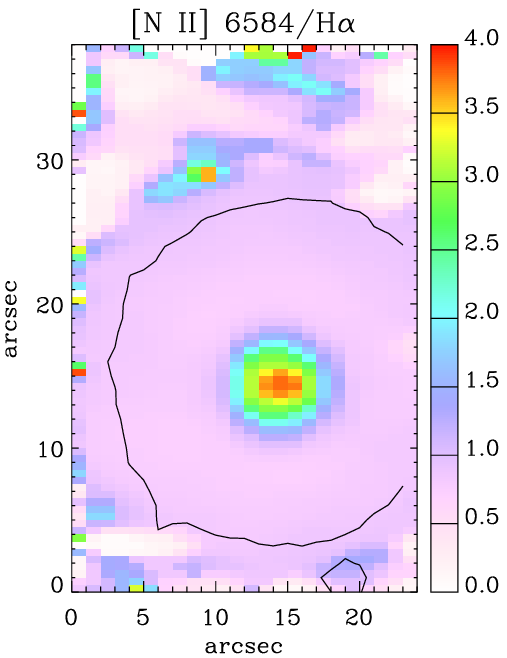}%
\includegraphics[width=1.1in]{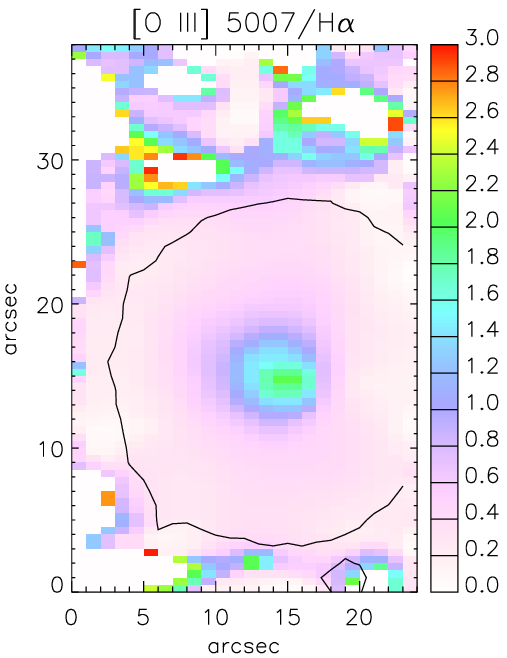}\\
\caption{From left to right, spatial distribution maps of flux intensity, continuum, LSR radial velocity and velocity dispersion of H$\alpha$ $\lambda$6563 emission line profile; flux ratio maps of $[$N~{\sc ii}$]$ $\lambda$6584 and $[$O~{\sc iii}$]$ $\lambda$5007 to H$\alpha$ $\lambda$6563 emission line, respectively for (a) M\,3-30, (b) Hb\,4, (c) IC\,1297, (d) Th\,2-A and (e) K\,2-16. Flux unit is in $10^{-15}$~erg\,s${}^{-1}$\,cm${}^{-2}$\,spaxel${}^{-1}$, continuum in $10^{-15}$~erg\,s${}^{-1}$\,cm${}^{-2}$\,{\AA}${}^{-1}$\,spaxel${}^{-1}$, and velocity unit in km\,s${}^{-1}$. North is up and east is toward the left-hand side. 
}
\label{wc1:ifu_map}%
\end{center}
\end{figure*}

\section{Observatios\label{wc1:sec:kinematic}}

The optical integral field unit (IFU) spectra of PNe analyzed in this study \citep{Danehkar2013b} were obtained at Siding Spring Observatory, Australia, using the Wide Field Spectrograph \citep[WiFeS;][]{Dopita2010} on the ANU 2.3 m telescope. WiFeS is an image-slicing IFU feeding a double-beam spectrograph. It has a field-of-view of $25\arcsec\times38\arcsec$  and a spatial resolution of  $1\arcsec$.  Our observations were carried out with the R7000 grating ($R\sim 7000$). We acquired series of bias, dome flat-field frames, twilight sky flats, arc lamp exposures, wire frames, spectrophotometric standard stars for flat-fielding, wavelength calibration, spatial calibration and flux calibration \citep[described in detail by][]{Danehkar2013a,Danehkar2014}. 

For each spatially resolved emission-line profile, we extracted flux intensity, continuum, radial velocity, and velocity dispersion. Each emission-line profile for each spaxel is fitted to a single Gaussian curve using the IDL-based routine \textsc{mpfit} \citep{Markwardt2009}.

\begin{figure}
\begin{center}
{\scriptsize (a) M\,3-30}\\ 
\includegraphics[width=2.5in]{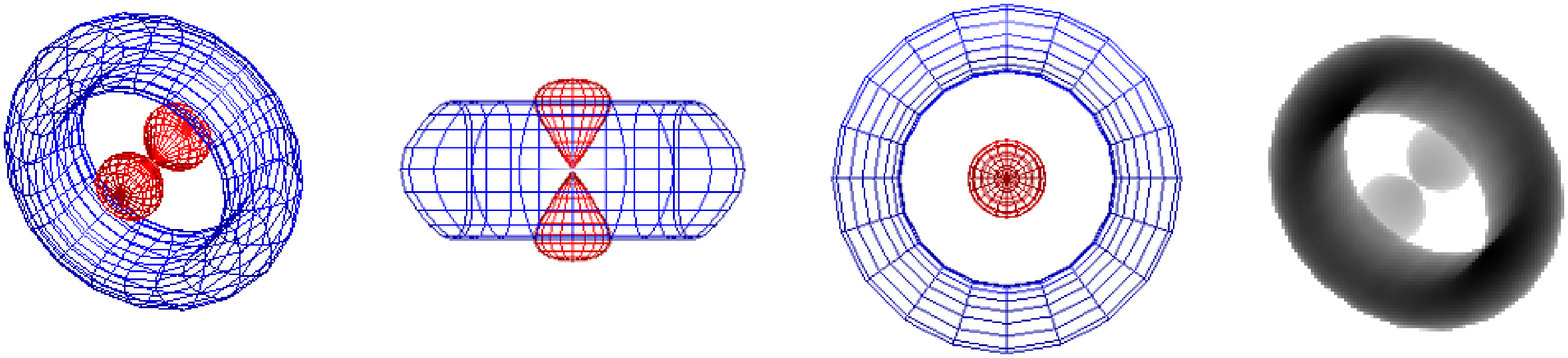}\\
{\scriptsize (b) Hb\,4}\\ 
\includegraphics[width=2.5in]{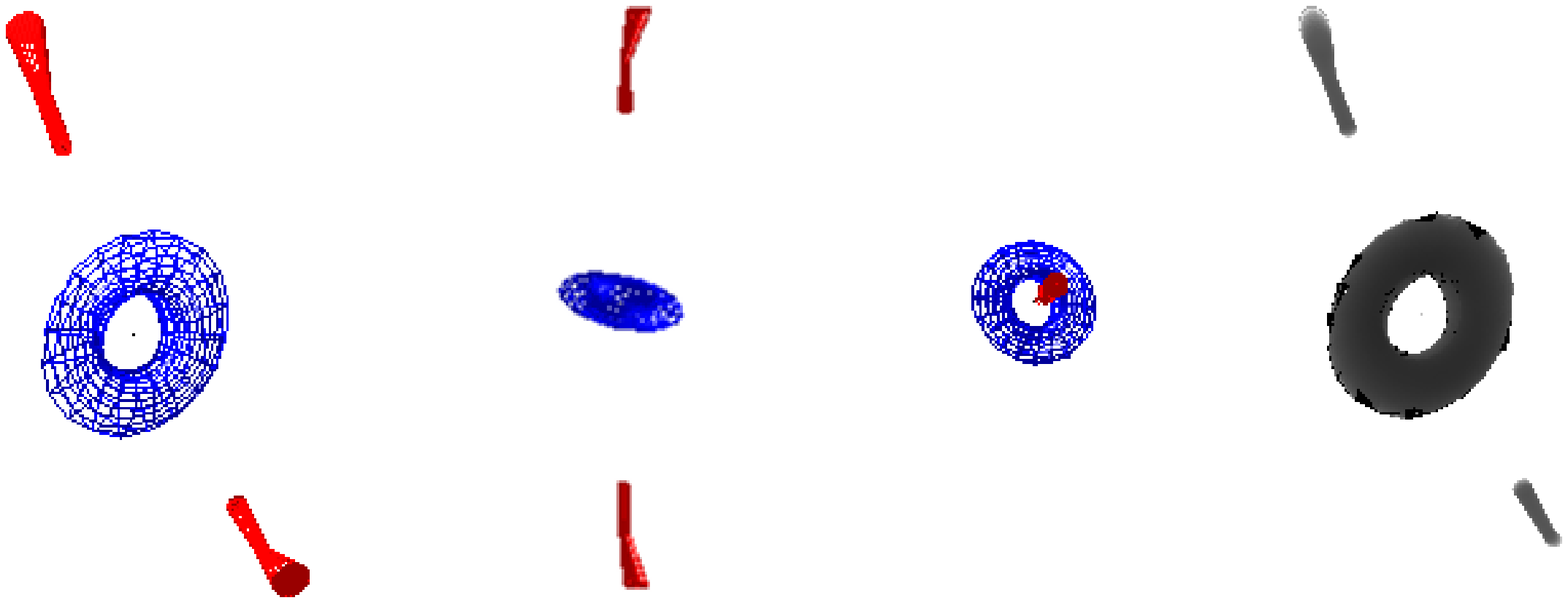}\\
{\scriptsize (c) IC\,1297}\\ 
\includegraphics[width=2.5in]{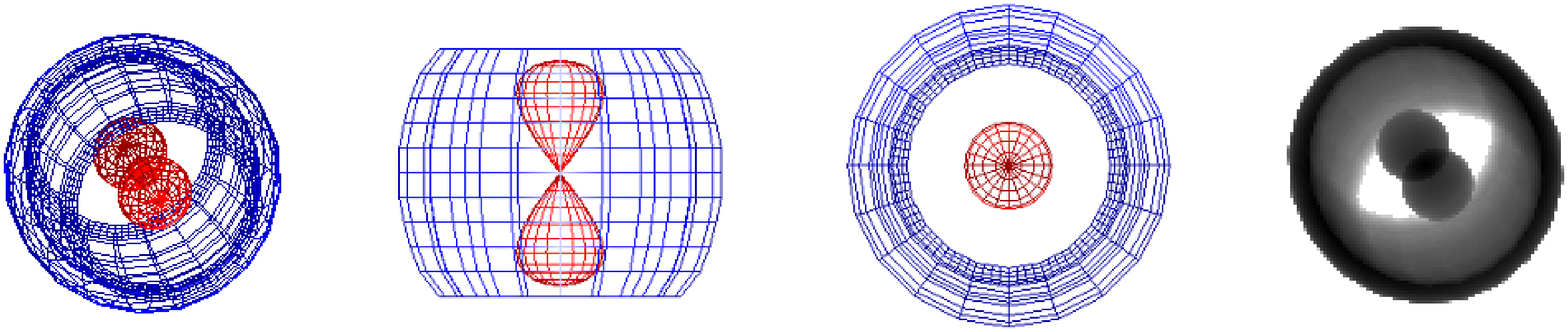}\\
{\scriptsize (d) Th\,2-A}\\ 
\includegraphics[width=2.5in]{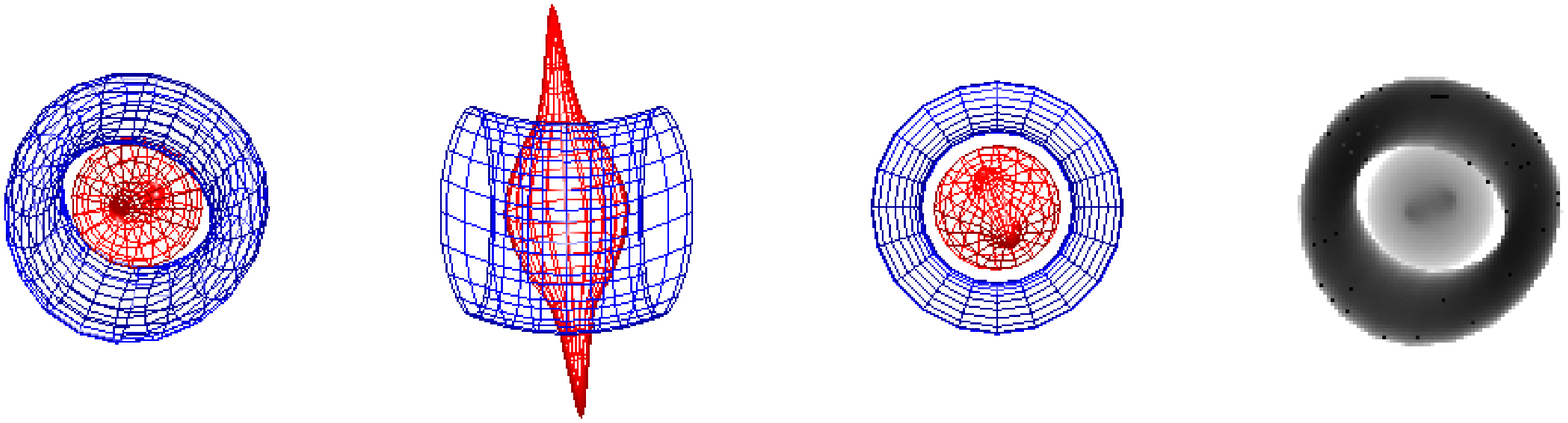}\\
{\scriptsize (e) K\,2-16}\\ 
\includegraphics[width=2.5in]{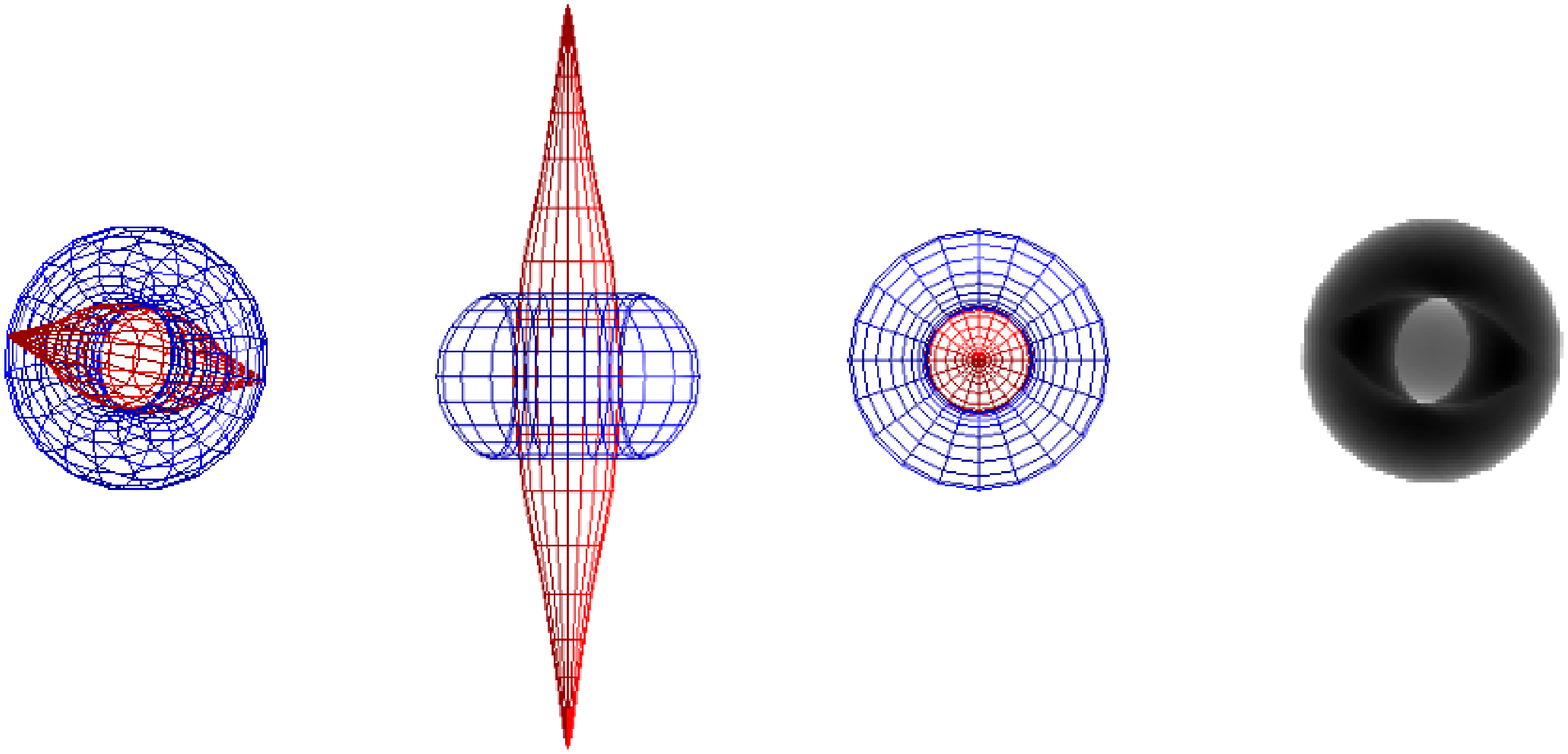}
\caption{\textsc{shape} mesh model before rendering at the best-fitting inclination, two different orientations (inclination: 90$^{\circ}$ and 0$^{\circ}$, respectively), and corresponding rendered model for (a) M\,3-30, (b) Hb\,4, (c) IC\,1297, (d) Th\,2-A and (e) K\,2-16. 
}
\label{wc1:shap}%
\end{center}
\end{figure}

Figure~\ref{wc1:ifu_map} shows spatially resolved kinematics of M\,3-30, Hb\,4, IC\,1297, Th\,2-A and K\,2-16, which were produced by fitting a single Gaussian profile to the emission line H$\alpha$ for all spaxels across the IFU field. The emission line profile is resolved if its width is wider than the instrumental width. Contour lines in the figure depict the 2-D distribution of the H$\alpha$ emission obtained from the AAO/UKST SuperCOSMOS H$\alpha$ Survey \citep{Parker2005}, which can aid us in distinguishing the nebular borders. Figure~\ref{wc1:ifu_map} also shows flux ratio maps of $[$N~{\sc ii}$]$ $\lambda$6584 and $[$O~{\sc iii}$]$ $\lambda$5007 to H$\alpha$ $\lambda$6563 emission line, which can be used to identify different ionization zones, i.e. low-excitation and high-excitation regions. 

\section{Morpho-kinematic modeling\label{wc1:sec:morpho-kinematic}}

We have used the three-dimensional morpho-kinematic modeling program \textsc{shape} (version 4.5) described in detail by \citet{Steffen2006} and \citet{Steffen2011}. It uses interactively molded geometrical polygon meshes to generate the three-dimensional structure of gaseous nebulae. The program produces several outputs that can be directly compared with observations, namely the position-velocity diagram, velocity channels and appearance of the object on the sky. The program does not include explicit photo-ionization modeling, so the emissivity distribution for each spectral line is modeled on an ad-hoc basis of the observations of the corresponding emission line. The modeling procedure consists of defining the geometry (e.g. torus, sphere and cylinder and their modification), assigning an emissivity distribution and defining a velocity law as a function of position. For this paper, the velocity has been defined as radially outwards from the nebular center with a linear function of magnitude. The inclination and the model parameters are modified in an iterative process until the qualitatively fitting solution is produced.

\section{Comments on some objects\label{wc1:sec:discussion:individual}}

\textbf{M\,3-30}. A \textsc{shape} model for M\,3-30 was built with a torus and inner bi-conical structure for its two FLIERs. Previously, \citet{Stanghellini1993} also identified the morphology as elliptical with inner knots or filaments. From the model, we derived a position angle of $-44^{\circ} \pm 5^{\circ}$ and an inclination of $34^{\circ}\pm 5^{\circ}$ relative to the line of sight. The maximum observed velocity between the receding and approaching components of the torus corresponds to an expansion velocity of $\sim 27$ \,km\,s$^{-1}$ at the inclination of the PN. This expansion velocity is lower than the value of $\sim 36$ \,km\,s$^{-1}$ derived from the HWHM of the integrated line flux, which can be explained by two FLIERs embedded in the main body. Furthermore, we derived $v_{\rm jet}=\pm54$\,km\,s$^{-1}$ with respect to $v_{\rm sys}=67$\,km\,s$^{-1}$, from the maximum observed velocity measured in the bi-conical structure. 

\textbf{Hb\,4}. IFU maps show the presence of two FLIERs located furthest away from the ring shell. The \textit{HST} image shows that the ring shell is deformed and the FLIERs are diametrically aligned but are slightly bent, which could be because of interaction with the ISM. A \textsc{shape} model for Hb\,4 was constructed consisting of a ring and two point-symmetric knots. The HWHM measured from an aperture $10\times 10$ arcsec$^{2}$ located on the ring present the expansion velocity, $V_{\rm HWHM}=23$\,km\,s$^{-1}$, l.  
The maximum observed velocity between the receding and approaching FLIERs corresponds to the velocity of $v_{\rm jet}=\pm 150$ \,km\,s$^{-1}$ at the inclination of $40^{\circ}\pm 5^{\circ}$ and a position angle of 25$^{\circ} \pm 10^{\circ}$ with respect to the plane of the sky, in agreement with \citet{Lopez1997}.

\textbf{IC\,1297}. \citet{Stanghellini1993} classified M\,3-30 under the irregular nebulae. A morpho-kinematic model of this PN was built from an elongated cylinder. It has an inclination of $-160^{\circ}\pm 10^{\circ}$ relative to the line of sight. The IFU maps look nearly as a twin of M\,3-30. It has more likely a torus structure with two inner knots at a different inclination. From the \textsc{shape} model, we derived $v_{\rm exp}=26\pm5$\,km\,s$^{-1}$. Our IFU kinematic maps suggests it may have a pair of knots embedded in the main shell. 

\textbf{Th\,2-A}. The roughly rectangular shape of Th\,2-A may be related to interaction with the ISM. For the morpho-kinematic study, we adopted a torus surrounding a prolate ellipsoid (jets). The ring-like morphology of Th\,2-A is also visible in the H$\alpha$ image \citep{Gorny1999}. From the radial velocity map, the expansion velocity of the shell relative to the nebula center was found to be $v_{\rm exp}=34\pm 10$\,km\,s$^{-1}$ at the inclination of $-17^{\circ}\pm 4^{\circ}$ to the line of sight, in agreement with the HWHM velocity. Furthermore, the high velocity dispersion and large blueshift in the center are likely related to the FLIERs seen pole-on. The velocity dispersion in the center corresponds to a jet velocity of $v_{\rm jet}\gtrsim 100$ \,km\,s$^{-1}$. 

\textbf{K\,2-16}. K\,2-16 is seen as a faint, circular nebula \citep{Schwarz1992}. However, the IFU observations suggest that this PN cannot have an spherical structure, and indeed has an aspherical morphology, either elliptical or bipolar. The \textsc{shape} model of K\,2-16 includes an elliptical structure surrounding a thin prolate ellipsoid. Taking the inclination of $20^{\circ}$ found by the best-fitting morpho-kinematic model, we derived a velocity of $v_{\rm jet}=\pm58$\,km\,s$^{-1}$ from the difference between velocity components in the IFU maps.

\section{Discussion\label{wc1:sec:discussions}}

Our aim in this work was to determine morphological features of Galactic PNe surrounding WR-type stars and \textit{wels}. The capabilities of the IFU observations have allowed us to identify their kinematic structures. The overall results indicates that they have axisymmetric morphologies. Some of them show elliptical shapes with FLIERS such as M\,3-30 and Hb\,4. Recently, \citet{Akras2012} also identified other WR PNe having axisymmetric shapes and fast bipolar outflows. Moreover, the recent \textit{Chandra X-Ray Observatory} survey of PNe \citep{Kastner2012} indicates that most elliptical PNe with FLIERs display ``diffuse X-ray'' sources.  Diffuse X-ray sources and ``hot bubbles'' were also found in some aspherical PNe around WR stars, e.g. NGC\,40 \citep{Montez2005} and NGC\,5315 \citep{Kastner2008}, which could be evidence for collimated jets and wind-wind shocks. The ``hard X-ray'' emission may suggest the presence of a binary companion, whereas soft, diffuse X-ray emission is more likely related to shocks \citep{Kastner2012}. \citet{Nordhaus2006} suggested that a binary system consisting of a low-mass star ($<0.3{\rm M}_{\odot}$) and an AGB star undergoes a common envelope phase, which can lead to binary-induced equatorial outflow and nebular aspherical morphology. While no binary companion has been found in the WR PNe of our sample, the merger scenario of a low-mass companion with an AGB star during the AGB phase may explain their typical aspherical morphologies. 

We believe future observations of PNe with  WR-type stars are necessary to investigate possible mechanisms, which shape their aspherical morphologies. 
The presence of binary companions should be inspected. However, it is extremely difficult to detect a low-mass star or a planet, which could also influence the shaping of its nebula. 
In-depth studies of their central engine are required to unravel the puzzle of the mechanism responsible for their morphologies, which as yet remains unanswered.

%%% ACKNOWLEDGMENTS (IF ANY) %%%%%%%%%%%%%%%%%%%%%%%%%%%%%%%%%%%%%%%%

\acknowledgments

AD acknowledges the award of an MQ Research Excellence Scholarship, and travel supports from Astronomical Society of Australia, IAU and Australian Institute of Physics. WS acknowledges support from grant UNAM-PAPIIT 101014. QAP acknowledges support from Macquarie University and AAO. 

% from the Organizing Committee of the 12th Asia-Pacific Regional IAU Meeting (APRIM 2014).

%%% APPENDICES (IF ANY) %%%%%%%%%%%%%%%%%%%%%%%%%%%%%%%%%%%%%%%%%%%%%

%%% CALL LIST OF REFERENCES (natbib STYLE) %%%%%%%%%%%%%%%%%%%%%%%%%%

\end{document}